\documentclass[a4paper,11pt]{article}
\usepackage{graphics,graphicx,epsfig,wrapfig}
\usepackage{longtable}  
\usepackage{amsmath,verbatim,amssymb,color,lscape}
\usepackage{kotex}
\usepackage{natbib}
\usepackage{lastpage}
\usepackage{multirow} 
\usepackage{authblk}
\usepackage{bm}
\usepackage{indentfirst}
\usepackage[ruled,vlined]{algorithm2e}
\usepackage{algorithmic}
\usepackage[normalem]{ulem}
\usepackage[export]{adjustbox}
\usepackage{url}
\usepackage{booktabs}
\usepackage{array}
\usepackage{makecell}
\usepackage{subcaption}
\usepackage{xcolor}

\newcolumntype{P}[1]{>{\raggedright\arraybackslash}p{#1}}

\newcommand{\hide}[1]{}

\makeatletter

\usepackage[outerbars,color]{changebar}
\ifx\pdfoutput\undefined
\else\ifnum\pdfoutput>0
  \usepackage{pdfcolmk}
\fi\fi
\cbcolor{black}

\usepackage[margin=1.0in]{geometry}
\usepackage{tikz}
\usetikzlibrary{bayesnet}
\usetikzlibrary{fit,positioning}

\newfont{\rmm}{cmr10 at 11pt}
\rmm

\title{Analyzing Process Data from Computer-Based Assessments: A Tutorial on Preprocessing, Feature Extraction, and Model-Based Inference}

\author[1]{Daeun Hwangbo}
\author[1]{Junyeong Park}
\author[3]{Minjeong Jeon}
\author[1,2]{Ick Hoon Jin}
\affil[1]{Department of Statistics and Data Science, Yonsei University. Republic of Korea.}
\affil[2]{Department of Applied Statistics, Yonsei University. Republic of Korea.}
\affil[3]{School of Education and Information Studies, University of California, Los Angeles. USA.}

\date{}

\begin{document}
\maketitle

\begin{abstract}
Computer-based assessments routinely generate detailed interaction logs---commonly referred to as process data---that record every action a respondent performs during task completion, yet systematic preprocessing guidance, integrated analytical workflows, and cross-method consistency checks remain scarce in the literature. This paper provides a unified, end-to-end analytical framework for analyzing process data from large-scale assessments --- covering the full pipeline from raw log preprocessing to model-based inference--- using the Programme for the International Assessment of Adult Competencies (PIAAC) Problem Solving in Technology-Rich Environments (PS-TRE) domain as an illustrative example. We first present a systematic preprocessing pipeline---including timestamp correction, duplicate removal, action block consolidation, and LLM-assisted standardization---that transforms raw event-level logs into analysis-ready action sequences. We then review and demonstrate two complementary families of analytical methods. The first consists of feature-based methods and their downstream applications, including descriptive process indicators, n-gram analysis with TF--IDF weighting, multidimensional scaling, and process data-informed differential item functioning (DIF) analysis. The second consists of model-based approaches, namely hidden Markov models and the subtask identification procedure. Empirical illustrations using the United States sample illustrate that n-gram-based behavioral clusters carry differential diagnostic information primarily among incorrect respondents, that multidimensional scaling-derived features comprehensively reconstruct observed behavioral variables, and that process-informed DIF analyses can identify and mitigate construct-irrelevant sources of group differences. Reproducible \path{R} code implementations are provided for all major techniques.
\end{abstract}


\section{Introduction}\label{sec:intro}

The increasing adoption of computer-based assessments in educational and psychological measurement has substantially changed how we understand human problem-solving behavior. Unlike traditional paper-based tests that capture only final responses, digital assessments automatically record detailed interaction logs---commonly referred to as \emph{process data} or \emph{log data}---that trace every action a respondent performs during task completion \citep{greiff2016understanding, kroehne2018conceptualize, goldhammer2021byproduct, maddox2023uses, chen2024understanding}. These behavioral records reveal not only what answers respondents produce but also how they arrive at those answers, thereby uncovering cognitive strategies, problem-solving patterns, and sources of difficulty that remain invisible when only final outcomes are examined. This richer view of test-taking behavior has opened new research directions in understanding individual differences, improving predictive models of performance, and addressing fairness concerns in educational measurement \citep{stadler2020first, eichmann2020exploring, he2021leveraging, reis2021improving, hahnel2023patterns}. Throughout this paper, we use \emph{process data} as an umbrella term for time-stamped interaction logs (log files) generated by the assessment platform.

Despite the growing recognition of process data's value, researchers face substantial challenges when attempting to analyze these complex behavioral records. Raw log files often originate from hierarchical formats, mix meaningful user actions with system-generated events, and require extensive preprocessing before any meaningful analysis can begin \citep{liu2023turning, qin2025preprocessing}. Preprocessing decisions---such as how to handle timestamp irregularities, remove redundant events, and consolidate multi-part actions---can substantially influence downstream analyses, yet standardized pipelines remain limited. Furthermore, the diversity of available analytical methods presents considerable difficulty for practitioners. Approaches range from simple aggregate indicators (e.g., time on task, number of actions) to sophisticated statistical models (e.g., hidden Markov models, multidimensional scaling), each carrying different assumptions, strengths, and interpretive considerations. More importantly, the literature still has several gaps. First, although preprocessing decisions can strongly affect downstream results, systematic evaluations of preprocessing workflows remain limited. Second, existing studies often focus on a single analytical approach, making it difficult for applied researchers to see how raw process data are converted into analysis-ready action sequences and subsequently used for substantive analysis. Third, few studies examine whether similar substantive conclusions emerge across different analytical approaches, even though such consistency is important for building confidence in behavioral interpretations.

This paper addresses these challenges by providing a unified, end-to-end analytical framework for analyzing process data from large-scale assessments, using the Programme for the International Assessment of Adult Competencies (PIAAC) as an illustrative example. PIAAC is well suited to this purpose. As a widely used international assessment with publicly available data, it systematically collects high-resolution behavioral logs and has been the focus of numerous methodological innovations in recent years \citep{he2015identifying, he2016analyzing, tang2020latent, ulitzsch2023machine, zhang2024external, liu2025uncovering, li2025exploring}. By focusing on PIAAC's Problem Solving in Technology-Rich Environments (PS-TRE) domain, we illustrate each method with real data while highlighting the item-specific considerations that arise when interpreting process data in realistic, complex task environments.

Our objectives are threefold, each corresponding to one of the gaps identified above. First, we provide systematic guidance on preprocessing raw log files by detailing common issues and presenting concrete strategies for addressing them. We also show how LLM-assisted workflows can accelerate implementation while still requiring principled validation. Second, we present preprocessing and analysis as parts of a unified pipeline, demonstrating how raw process data are transformed into analysis-ready action sequences and then used across a range of substantive analytical approaches. Third, we provide empirical illustrations based on a common dataset and item, enabling direct comparison of conclusions drawn from different analytical families. To support reproducibility and accessibility, we supply \path{R} code implementations for all major techniques.

The remainder of this paper follows the progression from raw process data to substantive analysis. Section~\ref{sec:piaac} introduces PIAAC process data, describing the structure of the log files and the characteristics of the assessment items that generate them. Section~\ref{sec:preprocess} addresses preprocessing, providing step-by-step procedures to transform raw logs into analysis-ready action sequences. Section~\ref{sec:notation} introduces the notational framework used consistently throughout the remainder of the paper. Section~\ref{sec:method_feature} presents feature-based methods and their downstream applications, while Section~\ref{sec:method_advance} introduces model-based approaches for studying latent problem-solving processes and strategies. Section~\ref{sec:conclusion} concludes with a discussion of methodological considerations, practical recommendations, and directions for future research.

\section{PIAAC and Process Data}\label{sec:piaac}

\subsection{The PIAAC Assessment and the PS-TRE Domain}\label{subsec:piaac_pstre}

The Programme for the International Assessment of Adult Competencies (PIAAC) is a large-scale international assessment program developed and coordinated by the Organisation for Economic Co-operation and Development \citep{oecd2012literacy}. PIAAC's flagship study---the Survey of Adult Skills---is a computer-based household survey of adults aged 16--65 that measures proficiency in key information-processing skills regarded as essential for full participation in modern societies and economies \citep{desjardins2013oecd}. The first cycle of PIAAC was administered in three rounds between 2011 and 2018 across 39 participating countries, and the second cycle was launched in 2022 with 31 participating countries \citep{oecd2024piaacdata}.

PIAAC assesses three core domains of adult skills. First, \emph{literacy} refers to the capacity to understand, evaluate, use, and engage with written texts in order to participate in society, achieve personal goals, and develop knowledge and potential \citep{piaac2009literacy}. Second, \emph{numeracy} involves accessing, using, interpreting, and communicating mathematical information and ideas to manage mathematical demands in a range of adult-life contexts \citep{piaac2009numeracy}. Third, \emph{problem solving in technology-rich environments} (PS-TRE) assesses the ability to use digital technology, communication tools, and networks to acquire and evaluate information, communicate with others, and perform practical tasks \citep{piaac2009pstre}.

Among these three domains, PS-TRE is of particular relevance to process data research because it directly targets adults' digital problem-solving capabilities in a standardized, large-scale setting. PS-TRE tasks are delivered exclusively through computer-based assessment, requiring respondents to interact with simulated digital environments---such as email clients, web browsers, spreadsheets, and word processors---to complete realistic problem-solving scenarios \citep{oecd2012literacy, chung2015adults}. Tasks vary considerably in complexity: simpler items require respondents to perform straightforward, single-application operations, whereas more complex items demand navigation across multiple applications, evaluation of information from multiple sources, and integration of findings into a coherent solution. For example, the Party Invitations item (U01a) involves a relatively straightforward email-management task in which respondents sort received messages into pre-existing folders according to an explicit criterion. In contrast, the Lamp Return item (U23) requires respondents to navigate a multi-page retailer website, complete an online exchange form, and confirm the transaction---a multi-step process that demands flexible strategy use and sustained attention across application contexts.

\subsection{Process Data in PIAAC: Characteristics and Challenges}\label{subsec:piaac_process}

A key feature of PIAAC's computer-based assessment is the systematic collection of process data through detailed log files. In the PS-TRE domain, process data comprise timestamped records of respondents' interactions with the testing application, capturing each discrete action---such as mouse clicks, page navigation, and form submissions---as it occurs during task completion. Unlike traditional outcome measures that record only whether a response is correct or incorrect, these behavioral records provide a fine-grained trace of how individuals approach and execute problem-solving tasks in technology-rich environments \citep{organisation2019beyond, goldhammer2020analysing}. The systematic collection of such logs has made PIAAC process data an important resource for methodological research on behavioral analysis in large-scale educational assessments \citep{schleicher2008piaac, vera2017beyond, hamalainen2019makes}. A key advantage of this richness is that process data can surface behavioral differences that remain entirely invisible in final scores. Consider two respondents who produce the same incorrect final answer: one may have followed a systematic but ultimately misdirected strategy, while the other may have engaged in unstructured exploration with minimal effort. Such differences are undetectable from outcome scores, yet are directly observable in action sequences.

Beyond this diagnostic capacity, process data offer several broader advantages for understanding test-taking behavior. First, they enable the examination of dynamic behavioral patterns over time by capturing not only which actions occur, but also when they occur and in what order \citep{he2015identifying, he2016analyzing, ulitzsch2023machine, zhou2024investigating}. Second, they facilitate the inference of problem-solving strategies by identifying characteristic behavioral sequences associated with successful or unsuccessful performance \citep{tang2020latent, tang2021exploratory, xiao2021exploring, zhang2024external, liu2025uncovering}. Third, they support behavioral clustering, enabling researchers to investigate how distinct process patterns relate to proficiency estimates and other outcome measures \citep{he2019clustering, wang2023subtask}.

Despite these advantages, the analysis of process data presents substantial challenges. A primary difficulty is that log files often contain noise or irrelevant events, as not all recorded actions are equally informative about underlying cognitive processes \citep{he2015identifying, he2016analyzing}. A second challenge concerns the inconsistency of event definitions across assessment platforms and items \citep{chen2024understanding}: different tasks may employ different logging schemes, and similar actions may be recorded under varying labels or at different levels of granularity, complicating cross-task and cross-national comparisons. A third challenge is the absence of standardized preprocessing procedures. Unlike traditional survey data, for which well-established pipelines exist, process data require a series of careful decisions about data quality screening, event selection, sequence construction, and feature engineering---decisions that can substantially influence downstream analyses and conclusions \citep{goldhammer2020analysing}.

These characteristics—noise-laden logs, inconsistent event definitions, and the absence of standardized preprocessing procedures—directly motivate the analytical framework developed in this paper. To provide context for that framework, we first describe the 14 PS-TRE items and their task characteristics (Section~\ref{subsec:pstre_item}) before examining the structure of the raw log files through a worked example (Section~\ref{subsec:pstre_rawlog}).

\subsection{Item Context in PIAAC PS-TRE Domain}\label{subsec:pstre_item}


The PIAAC PS-TRE assessment comprises 14 computer-based items designed to measure adults' problem-solving skills in technology-rich environments \citep{oecd2012literacy, piaac2009pstre}. These items span a broad range of difficulty levels and differ substantially in both cognitive demands (e.g., information evaluation and integration) and technical demands (e.g., navigation, tool use, and switching between applications).

Table~\ref{tab:pstre_items} summarizes the 14 PS-TRE items, including item identifiers, names, and brief task descriptions. These items exhibit considerable variation in both their structural characteristics and interaction demands, supporting a comprehensive assessment of problem-solving competencies and generating diverse process data that reflect varied behavioral patterns.

\begin{longtable}{@{} l P{0.26\linewidth} P{0.62\linewidth} @{}}
\caption{
Summary of the 14 PS-TRE items in the PIAAC assessment, including item identifiers, names, and brief task descriptions.
}
\label{tab:pstre_items} \\
\toprule
ID & Item Name & Task Description \\
\midrule
\endfirsthead

\toprule
ID & Item Name & Task Description \\
\midrule
\endhead

\midrule
\endfoot

\bottomrule
\endlastfoot

U01a & Party Invitations -- Can/Cannot Come & review invitation response emails and categorize them into existing folders based on whether each invitee can attend or cannot attend \\

U01b & Party Invitations -- Accommodations & create and use email folders to categorize invitation responses, identifying messages with special accommodation requests based on an explicit criterion and filing them into the appropriate folder \\

U03a & CD Tally & organize a large amount of information in a multi-column spreadsheet, apply an explicit criterion, and determine the required value to report as the outcome \\

U06a & Sprained Ankle -- Site Evaluation Table & evaluate several websites listed on a search-engine results page by assigning each site a single label based solely on explicit credibility criteria \\

U06b & Sprained Ankle -- Reliable/Trustworthy Site & navigate across multiple web pages to apply explicit credibility criteria and select the single most reliable and trustworthy website for treating a sprained ankle \\

U21 & Tickets & use an online ticketing application, switching between web pages and tools to find games with available tickets and reserve seats for all games that satisfy multiple explicit constraints for a group \\

U04a & Class Attendance & use information provided in an email to populate and organize a spreadsheet, applying the stated criteria to compute and report total participation for each training class by department \\

U19a & Club Membership -- Member ID & locate a specific member (Steven Lopez) in a membership spreadsheet using an explicit criterion and email the requested ID number to the club secretary \\

U19b & Club Membership -- Eligibility for Club President & use a multi-column membership spreadsheet to identify and mark members who meet multiple explicit eligibility criteria for club president, drawing the criteria from the accompanying club newsletter/note \\

U07 & Book Order & navigate across multiple web pages to identify the book option that best matches a set of given constraints and complete the selection by clicking the purchase option on the chosen page \\

U02 & Meeting Room & review email requests for meeting rooms and use a simulated online reservation system to schedule as many requests as possible for a given date while managing multiple constraints across applications \\

U16 & Reply All & compose a reply email that includes specified information and ensure it is sent to multiple intended recipients \\

U11b & Locate Email -- File 3 E-mails & organize several emails by inferring the appropriate destination folders from the subject header and message content \\

U23 & Lamp Return & navigate a retailer website to arrange an exchange for an incorrectly delivered lamp, completing a specified return/exchange transaction by submitting a request, filling out an online form, and confirming the order \\
\bottomrule
\end{longtable}

Throughout this paper, all empirical illustrations and analyses are based on the U.S. sample of the PIAAC PS-TRE data, unless otherwise noted. This choice reflects both the availability of publicly accessible data and the prominence of the U.S. sample in the existing methodological literature on PIAAC process data.

To illustrate the heterogeneity of behavioral demands across items, Table~\ref{tab:pstre_stats} presents item-level performance and process indicators for the U.S. sample, including accuracy, mean item scores, the median number of recorded events, and median time on task. As shown, there is substantial variation in performance across items. For dichotomous items, accuracy rates span a wide range, reflecting marked differences in item difficulty. For polytomous items scored on a 0--3 scale, mean scores are generally low, suggesting that many respondents received only partial credit. Process-level indicators exhibit similar heterogeneity: both the median number of recorded events and the median time on task vary considerably across items. Taken together, these patterns underscore the diversity of cognitive and behavioral demands across PS-TRE items and motivate the use of process data to characterize individual differences in problem-solving strategies.

\begin{table}[htb]
\centering
\caption{
Item-level performance and process indicators for the 14 PS-TRE items (United States sample). Accuracy denotes the proportion of respondents answering correctly for dichotomous items and the proportion achieving the maximum score (3) for polytomous items. Score reports the mean item score for polytomous items only (denoted by --- for dichotomous items). \# of Events and Time (seconds) are item-level medians across respondents. $^\dagger$~Polytomous item (0--3 scale).
}
\label{tab:pstre_stats}
\small
\begin{tabular}{l l c c c c c }
\toprule
ID & Item Name & \# of respondents & Accuracy & Score & \# of Events & Time (seconds)\\
\midrule
U01a$^\dagger$ & Party Invitations -- 1 & 1340 & 0.504 & 1.849 & 33 & 99.0 \\
U01b & Party Invitations -- 2 & 1337 & 0.434 & --- & 67 & 138.1 \\
U03a & CD Tally & 1335 & 0.367 & --- & 26 & 102.9 \\
U06a & Sprained Ankle -- 1 & 1332 & 0.268 & --- & 18 & 128.9 \\
U06b & Sprained Ankle -- 2 & 1331 & 0.498 & --- & 41 & 91.2 \\
U21 & Tickets & 1331 & 0.385 & --- & 41 & 177.9 \\
U04a$^\dagger$ & Class Attendance & 1331 & 0.099 & 0.395 & 73 & 231.8 \\
U19a & Club Membership -- 1 & 1341 & 0.659 & --- & 45 & 110.5 \\
U19b$^\dagger$ & Club Membership -- 2 & 1341 & 0.000 & 1.109 & 45 & 171.4 \\
U07 & Book Order & 1341 & 0.466 & --- & 43 & 101.1 \\
U02$^\dagger$ & Meeting Room & 1341 & 0.086 & 0.591 & 54 & 154.2 \\
U16 & Reply All & 1341 & 0.531 & --- & 78 & 112.3 \\
U11b$^\dagger$ & Locate Email & 1328 & 0.169 & 0.870 & 29 & 73.9 \\
U23$^\dagger$ & Lamp Return & 1328 & 0.297 & 1.127 & 43 & 82.9 \\
\bottomrule
\end{tabular}
\end{table}

Each item in the PS-TRE assessment constitutes a distinct digital environment with its own set of action possibilities and problem-solving demands; consequently, interpreting process data requires careful attention to item-specific contexts. Section~\ref{subsec:pstre_rawlog} illustrates this point through a detailed examination of item U01a, which serves as the running example used to motivate the preprocessing pipeline introduced in Section~\ref{sec:preprocess}.

\subsection{Digital Task Environment and Log-File Structure}\label{subsec:pstre_rawlog}

\begin{figure}[htb]
    \centering
    \includegraphics[width=0.85\linewidth]{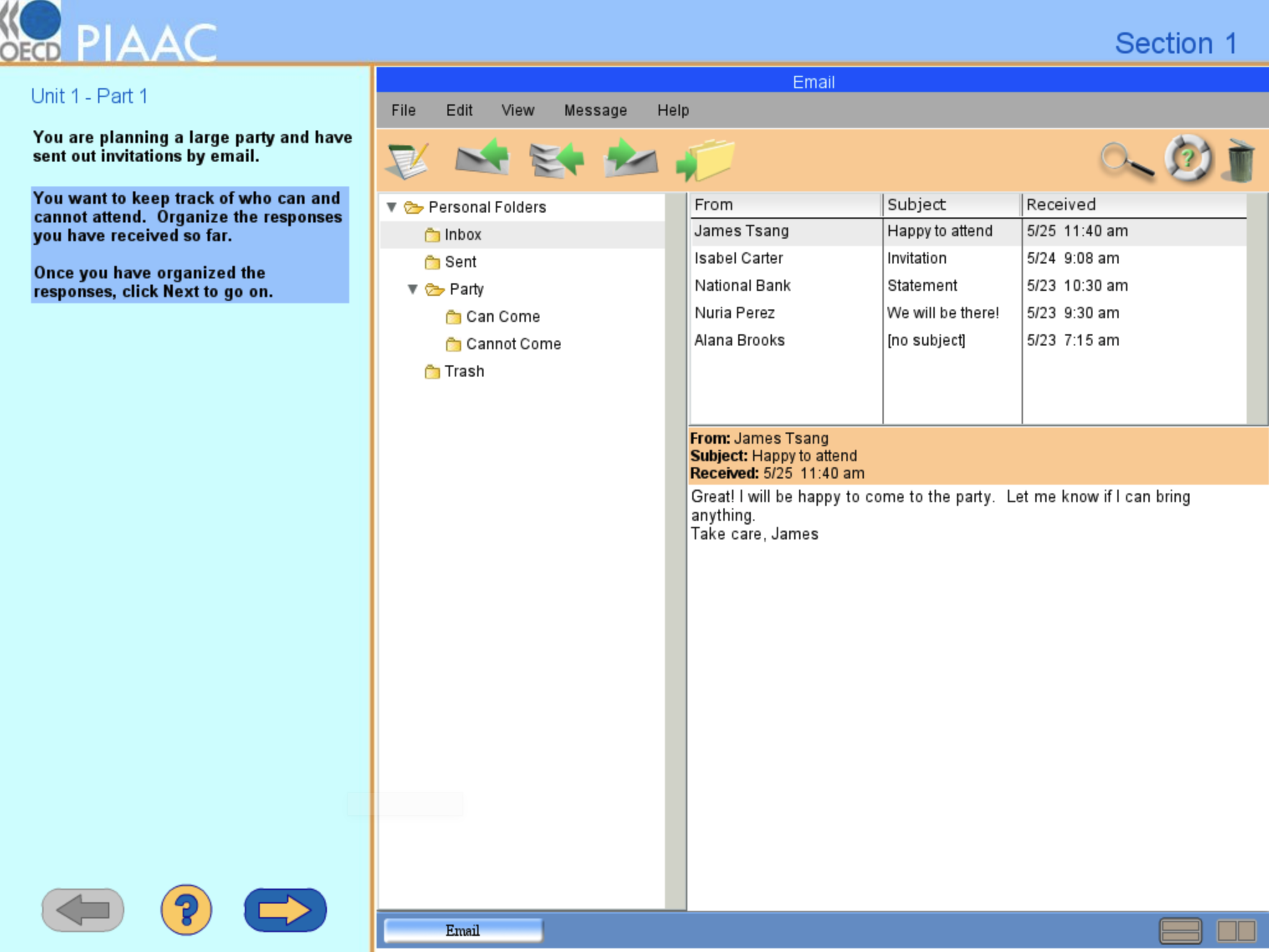}
    \caption{PIAAC PS-TRE task interface for the item \textit{Party Invitations -- Can/Cannot Come} (U01a). The left panel displays task instructions; the right panel shows the simulated email application with a folder tree, message list, and message preview pane.}
    \label{fig:digitaltask}
\end{figure}

To illustrate the digital environment in which PS-TRE tasks are administered and the resulting log-file structure, we use \textit{Party Invitations -- Can/Cannot Come} (U01a) as a running example throughout this paper. Figure~\ref{fig:digitaltask} shows the task interface for this item, which simulates an email client in which respondents must organize invitation responses into designated folders. The left panel displays the task instructions, informing respondents that they must track who can and cannot attend a party by sorting the received messages accordingly. The right panel displays the simulated email application, consisting of a folder tree in the left pane (e.g., Inbox, Sent, and Party subfolders) and a list of received emails in the upper-right pane; clicking an email renders its content in the lower-right pane. Although the interface appears relatively straightforward, successful completion requires multiple coordinated steps: reading each email, determining whether the sender can attend, and executing a drag-and-drop action to move the message into the appropriate folder. This combination of comprehension, decision-making, and interface manipulation is characteristic of PS-TRE tasks more broadly and generates rich, multi-step action sequences that are well suited to process data analysis.

\begin{figure}[htb]
    \centering
    \includegraphics[width=1\linewidth]{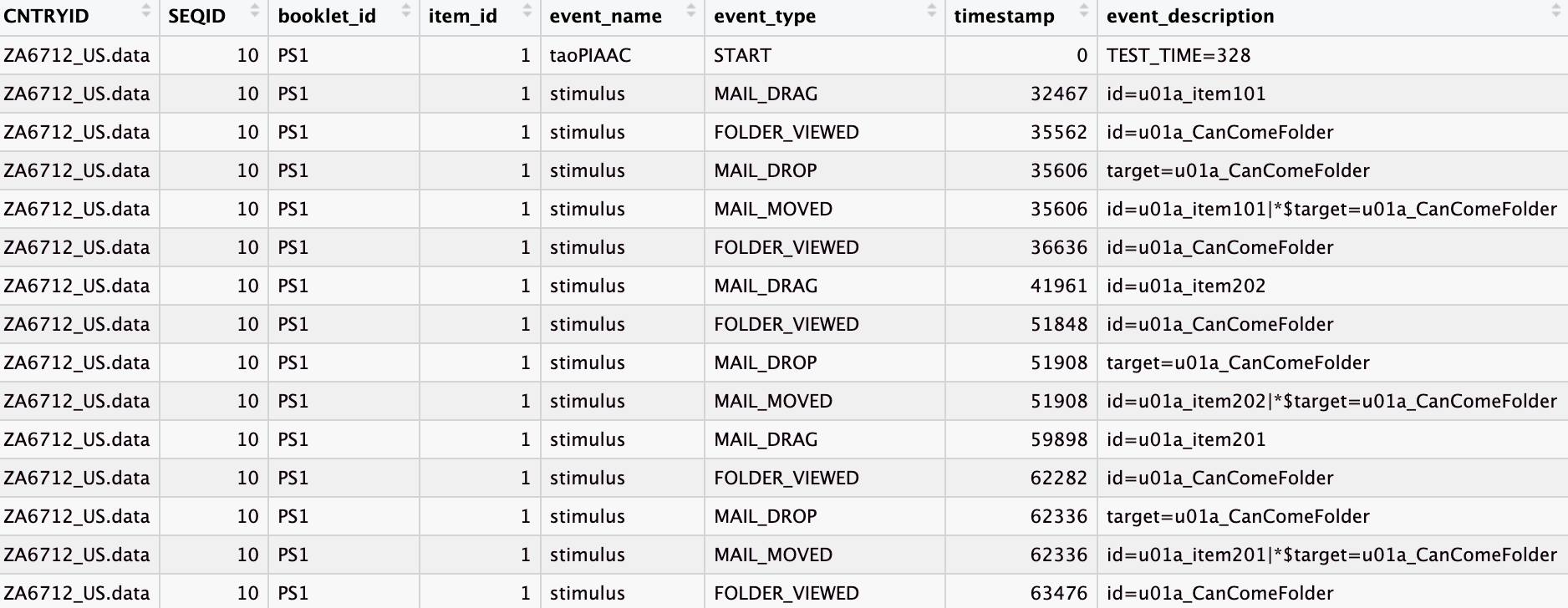}
    \caption{
    Excerpt of raw event-level process data for the item \textit{Party Invitations -- Can/Cannot Come} (U01a), extracted from the original XML format. Each row corresponds to one recorded event. Key fields include \texttt{event\_type} (the type of interaction), \texttt{event\_description} (contextual details of the specific object involved), and \texttt{timestamp} (elapsed time in milliseconds since item presentation).
    }
    \label{fig:rawlog}
\end{figure}

All respondent interactions with the interface are automatically captured as timestamped events in the assessment platform's log files. Figure~\ref{fig:rawlog} shows an excerpt of the raw log structure for U01a, obtained by parsing the original XML files into a flat, event-level table in which each row corresponds to one recorded event. Rather than storing only final responses, the platform records each discrete interaction separately, thereby enabling researchers to reconstruct the complete action sequence underlying task performance.

Each event record contains multiple fields that together describe what occurred and when. Three fields are particularly important for behavioral analysis. \path{event_name} identifies the source component that generated the event. \path{event_type} labels the type of interaction (e.g., clicking, dragging, or viewing). \path{event_description} provides contextual details about the specific object or action involved---for example, specifying which email was moved and to which folder. In addition, \path{timestamp}, measured in milliseconds elapsed since item presentation, captures the precise temporal ordering of actions, while other identifiers (country, respondent, booklet, and item IDs) link each event to the corresponding respondent and item.

For the item \textit{Party Invitations -- Can/Cannot Come} (U01a), \path{event_type} records specific email-management actions such as \path{MAIL_DRAG}, \path{FOLDER_VIEWED}, \path{MAIL_DROP}, and \path{MAIL_MOVED}, while \path{timestamp} preserves their sequential ordering. The \path{event_description} field distinguishes individual instances of the same action type---for example, identifying which email was moved and to which destination folder---thereby enabling the reconstruction of complete action sequences and the identification of behavioral patterns such as revisions or repeated attempts.

Taken together, Figures~\ref{fig:digitaltask} and~\ref{fig:rawlog} illustrate how PS-TRE tasks are delivered within realistic digital environments and how respondents' interactions are captured as granular, event-level logs. This combination of structured task interfaces and high-resolution behavioral recording yields rich process data for analyzing problem-solving strategies. However, the raw log files are not immediately suitable for analysis: they originate from hierarchical XML files, intermix respondent-initiated actions with system-generated events, and encode contextual information in semi-structured fields that vary across items and countries. Careful preprocessing is therefore required to parse and standardize records, filter uninformative events, and construct meaningful action sequences while preserving the temporal structure of the interaction stream.

Section~\ref{sec:preprocess} describes the preprocessing pipeline adopted in this paper, which addresses these challenges through data quality screening, event filtering, and sequence construction to produce analysis-ready action sequences.

\section{Data Preprocessing: From Raw Logs to Action Sequences}\label{sec:preprocess}

Preprocessing is a critical and often underappreciated step in process data analysis. Many studies on PIAAC process data provide little detail on how raw log events were cleaned and transformed prior to analysis, making it difficult for other researchers to reproduce or build upon their work. In PIAAC process data, a single user action may generate multiple event records, timestamps can become disordered due to system-level events such as restarts, and identical actions may be logged more than once. Because preprocessing decisions---such as how to handle timestamp irregularities, remove redundant events, and consolidate multi-part actions---can substantially influence downstream analytical outcomes, researchers must approach this stage with care. Although specific preprocessing strategies may vary depending on the research question and analytical method, certain procedures are broadly applicable across most frameworks and can be considered essential for PIAAC process data. This section describes common preprocessing issues and practical strategies for addressing them. Because implementing these preprocessing procedures often requires substantial scripting effort, this section also considers LLM-assisted preprocessing as a complementary support mechanism for researchers with limited programming experience.

\subsection{Common Preprocessing Issues}\label{subsec:issue}

Before any analytical model can be applied, raw log events must be transformed into a dataset in which each row represents a meaningful user action arranged in proper temporal order. This step is necessary because PIAAC log files often record low-level system events rather than directly providing the action units required for analysis. Among the many preprocessing challenges in PIAAC process data, several issues recur across items and should be addressed systematically before further analysis.

\subsubsection{Timestamp Reversal}
In some cases, log entries corresponding to earlier actions are recorded with later timestamps, disrupting the chronological ordering of actions within a sequence. To address this, log entries should be grouped by item and sequence identifier (SEQID) and sorted by timestamp within each group. A further complication arises from the system-generated restart event, which resets the timestamp to zero in the middle of a sequence. As a result, naively sorting all entries by timestamp would incorrectly place post-restart actions before actions that occurred earlier in real time. To preserve the true temporal order, each timestamp recorded after a restart should be adjusted by adding a restart-specific offset derived from the corrected timestamp assigned to the restart event. The restart action itself is assigned a corrected timestamp by adding half of the immediately following action's original timestamp to the preceding action's already-adjusted timestamp. This offset is then applied to all subsequent actions up to and including the next end event within the same SEQID, thereby preserving the internal temporal structure of the restarted segment. After this adjustment, the sequence can be interpreted as a single continuous timeline despite the temporary timestamp reset.

\subsubsection{Duplicate actions}
In some cases, multiple log entries with the same event type and timestamp appear within the same item and sequence. If left unaddressed, such repetitions can inflate action counts and distort the reconstructed sequence of user behavior. These duplicates can be resolved by grouping data by item and SEQID and removing exact duplicate events. An important exception involves keypress events, because repeated keystrokes may be logged with identical timestamps even when they reflect genuine user input rather than redundant system recording. In such cases, removing all repeated keypress records could eliminate behaviorally meaningful information about typing activity. For this reason, the decision to deduplicate keypress events should depend on the analytic purpose, particularly on whether repeated keystrokes are treated as substantive behavior or as irrelevant redundancy. Any such decision should be documented explicitly, because alternative deduplication rules may lead to different representations of user behavior.

\subsubsection{Multiple log entries per action}
The most complex preprocessing issue involves multiple log entries generated from a single user action. PIAAC raw logs record both core actions---the operations respondents actually perform---and ancillary actions that provide supplementary system details about those core actions. Preprocessing therefore requires identifying which entries should define the behavioral unit of analysis and which merely describe its execution context. One core action may be accompanied by several ancillary entries; for example, a respondent clicking a toolbar icon generates a \path{TOOLBAR} event (core) followed by a \path{DOACTION} event (ancillary) indicating which system function was activated. In many cases, the core entry contains most of the information that is analytically relevant for reconstructing user behavior. Accordingly, removing redundant ancillary entries can improve analytical efficiency while preserving the information needed for many action-based analyses. Two practical strategies can be used to address this issue.

The first is a \emph{time-threshold approach}, in which events recorded within a short interval after a core entry are grouped into a single action block. Because ancillary actions are recorded within a minimal time interval after the core action, a researcher-defined threshold can be used to group all events occurring within that window into a single action block. This approach requires no prior knowledge of specific core--ancillary pairings and is therefore straightforward to implement. Its main limitation is that the choice of threshold is inherently heuristic rather than being derived from an item-specific structural rule. As a result, the same threshold may perform differently across items or interaction contexts. This approach is therefore most appropriate when a coarse but operational preprocessing rule is sufficient for the intended analysis.

The second is a \emph{pattern-matching approach}, in which researchers inspect the log structure of the problem-solving process to identify recurrent core--ancillary combinations and collapse each combination into a single representative action. For example, if \path{TOOLBAR} is consistently followed by \path{DOACTION}, the researcher can define this sequence as a single action block and implement the rule programmatically. Although more time-intensive, this approach yields greater precision and allows researchers to define action blocks tailored to their specific analytical needs. Because each item type in the data examined here follows a consistent structural pattern in its core--ancillary pairings, preprocessing one representative item per type can serve as a template for all remaining items of the same type, substantially reducing the overall effort. In the present context, these item types correspond to three interface categories---Email, Spreadsheet, and Web---each of which tends to exhibit a common pairing structure across items. This makes it possible to balance preprocessing precision with practical feasibility in large-scale process data.

In addition to the three issues described above, analysts should consider whether certain system-generated markers warrant exclusion prior to analysis. The \path{start} event (recorded at $t = 0$) is a platform-generated item presentation marker rather than a behavioral interaction. Similarly, final navigation events---such as button clicks used to confirm task submission---may or may not constitute substantive problem-solving actions depending on the research question. The treatment of such boundary events should be explicitly documented, as it directly affects the computation of action-based indicators such as the number of actions and the time to first action.

\subsection{LLM-Assisted Preprocessing}\label{subsec:llm}

Implementing the preprocessing procedures described in the previous sections requires considerable programming proficiency and time investment. Translating the complex nested log structures of PIAAC data into analysis-ready formats often requires skills in data parsing, string manipulation, and iterative pattern matching that may not be central to the training of all process data researchers. As a result, the technical burden of preprocessing can become a practical barrier to the broader use of process data in empirical research. Recent large language models (LLMs) have shown strong potential for generating functional code from natural-language instructions, offering a more accessible implementation workflow for researchers with programming experience. Researchers can describe the desired preprocessing operations in a structured prompt that specifies the log format, the target transformation rules, and the expected output structure, and then obtain executable Python or R code in response. In this workflow, the researcher's role shifts from writing code to specifying requirements and verifying outputs---tasks that align more closely with their substantive expertise. We therefore view LLMs as assistive tools that lower the implementation cost of preprocessing rather than as substitutes for methodological decision-making. The usefulness of this approach, however, depends heavily on how clearly the prompt specifies the preprocessing task, the relevant log patterns, and the criteria for correct output. For this reason, the following subsections discuss prompt design principles and procedures for validating LLM-generated code.

\subsubsection{Prompt Design Principles}
The quality of code generated by LLMs depends strongly on how the task is specified in the prompt. A vague or underspecified prompt often leads to code that is syntactically valid but inconsistent with the intended preprocessing logic. Drawing on established principles for effective prompt design \citep{chen2025unleashing}, we highlight several strategies that were especially useful in generating preprocessing code for this study. The complete prompts used in this study are provided in Section~S1 of the Supplementary Materials.

\paragraph*{Role assignment}
Assigning the model a specific professional role at the outset helps orient its responses toward domain-appropriate solutions. In our prompt, the model was instructed to act as a senior data engineer with experience in educational assessment data analysis.
\paragraph*{Clear and precise instructions} Each operation should be described in concrete terms rather than left to the model's interpretation. For example, rather than instructing the model to correct timestamps in general terms, our prompt specified the exact formula for computing corrected restart timestamps and the range of rows to which the correction should be applied.
\paragraph*{Context provision} The prompt should supply sufficient background for the model to understand the data it will process. Our prompt included a description of the dataset, column definitions with data types, and sample rows from the raw log file.
\paragraph*{Structured formatting with delimiters} Distinct components of the prompt—such as data descriptions, processing steps, constraints, and examples—should be separated using headers, code blocks, or tables. This prevents the model from conflating instructions with data specifications.
\paragraph*{Task decomposition} Complex tasks should be broken into smaller, self-contained sub-tasks. Our prompt decomposed the full preprocessing pipeline into eight sequential steps, each to be implemented as a separate function, facilitating both code generation and subsequent debugging.
\paragraph*{Output format specification} The prompt should define the expected format of the generated output, including the programming language, permitted packages, coding conventions, and file structure. This reduces variability across generated responses and ensures compatibility with the analytical workflow used in the study.
\paragraph*{Few-shot prompting} Providing concrete input–output examples within the prompt constrains the model's interpretation more effectively than verbal descriptions alone. Our prompt included a sample input table and the corresponding expected output for each major preprocessing step, along with brief explanations of the transformation logic.
\paragraph*{Constraints and negative instructions} The prompt should state not only what the model should do but also what it must avoid. Our prompt explicitly prohibited in-place data modification, the use of unauthorized packages, and arbitrary deletion of missing values, all of which are patterns that LLMs may otherwise introduce by default.

Although discussed here in the context of PIAAC process data, these principles are broadly applicable to the use of LLMs in data preprocessing. The prompts developed for this study instantiate these principles in detail and, together with the generation environment metadata, are documented in Section~S1 of the Supplementary Materials.

\subsubsection{Validation of LLM-Generated Code}
To evaluate the validity of the LLM-based approach, we compared the preprocessing outputs generated by five LLMs against a manually written reference. The reference preprocessing scripts, developed and independently verified by the authors, were treated as the benchmark implementation. These scripts reflected the preprocessing logic described in the previous sections and were used to generate the expected output for each comparison. The same prompt, constructed according to the principles described above, was submitted to GPT-5.4, Claude Sonnet 4.6, Gemini 3.1 Flash-Lite, DeepSeek v3.2, and Qwen 3.5-27b \citep{openai2026gpt54, anthropic2026claudesonnet46, deepmind2026gemini31, liu2025deepseek, qwen35blog}. All models received the same task specification and were evaluated on the same data subset under the same execution criteria. For each model, both Python and R implementations were generated and executed on the same subset of PIAAC process data. The evaluation focused on two dimensions: executability and output accuracy. Executability assessed whether the generated code ran without errors on the target data. Output accuracy assessed the degree to which the resulting data frame matched the benchmark output, including the number of rows, column names, and cell-level values. Table \ref{tab:llm_comparison} summarizes the comparison results across the five models.

\begin{table}[ht]
\centering
\caption{
Comparison of LLM-generated preprocessing code against the benchmark output across five models and two programming languages. Accuracy reported after minimal manual correction.
}
\label{tab:llm_comparison}
\begin{tabular}{llcccc}
\hline
Model & Language & Executability & Accuracy (\%) & Issues \\
\hline
\multirow{2}{*}{GPT-5.4}          & Python & \checkmark &  99.99\% & - \\
                                  & R      & \checkmark & 99.85\% & - \\
\hline
\multirow{2}{*}{Claude Sonnet 4.6} & Python & \checkmark & 99.99\% & - \\
                                    & R      & \checkmark & 99.99\% & - \\
\hline
\multirow{2}{*}{Gemini 3.1 Flash-Lite}  & Python & \checkmark & 87.96\% & - \\
                                  & R      & \checkmark & 87.96\% & - \\
\hline
\multirow{2}{*}{Deepseek v3.2}  & Python & \checkmark & 99.99\% & - \\
                                  & R      & \checkmark & 99.99\% & - \\
\hline
\multirow{2}{*}{Qwen3.5-27b}  & Python & \checkmark & 95.33\% & f-string error \\
                                  & R      & \checkmark & 99.99\% & output path error \\
\hline
\end{tabular}
\end{table}

As shown in Table \ref{tab:llm_comparison}, most models produced immediately executable code with accuracy above 99\% in both Python and R. Gemini 3.1 Flash-Lite showed slightly lower accuracy in both languages; however, the discrepancies were limited to minor differences in timestamp values, while the sequence and types of actions were fully consistent with the benchmark output. Because timestamp deviations of this magnitude have negligible impact on downstream analyses, these outputs can still be considered practically adequate. Qwen 3.5-27b produced executable code in both languages, but the Python output contained an f-string formatting error that resulted in an incorrectly generated file, and the R output could not be saved at all because of a logic error. After correcting these issues through manual review, both scripts produced outputs consistent with the benchmark. These results indicate that well-designed prompts can elicit reliable preprocessing code from current LLMs, although systematic comparison against a reference output remains necessary to identify subtle discrepancies. The substantive outputs across models converged on the same analytical result despite being generated independently, supporting the validity of the prompt-based approach itself rather than reflecting the strength of any single model. The discrepancies that did arise were generally minor and identifiable through systematic comparison against the reference output.

\subsubsection{Considerations and Remarks}
Although the results support the practical usefulness of LLMs as an assistive tool for preprocessing, several considerations should be noted. Most importantly, effective prompt design presupposes that the researcher has conducted sufficient exploratory data analysis to understand the structure and characteristics of the raw data. Without this prior understanding, researchers cannot formulate prompts that are specific enough to elicit reliable code. Additionally, LLM-generated code should be treated as a draft rather than a final product. As observed in the validation results, outputs can appear plausible while containing subtle errors that are only detectable through careful verification against known reference cases. Human validation therefore remains an indispensable step in this workflow, regardless of which model is used. Researchers should also be aware that while LLM outputs may vary across model versions and runs due to their inherently stochastic nature, the downstream task outcomes derived from these outputs remain consistent and reproducible.

While the present results indicate that downstream outcomes were reproducible across models, this is not guaranteed in general and depends on how prompts are designed and maintained over time. Because LLMs are updated frequently and generate code stochastically, prompts should be designed for robustness across models rather than optimized for a single model version. We recommend first constructing a baseline prompt grounded in established prompt-design principles, such as the explicit specification of processing steps, input-output examples, and verification criteria, and then tuning this baseline to the particular model in use. To support reproducibility over time, researchers should document the exact prompt text together with the model name, version, and date of use. Furthermore, when a model is updated, the outputs of a previously reliable prompt may change; researchers should therefore re-verify that the regenerated code still reproduces the expected reference results before reuse.

Additionally, data confidentiality should be carefully considered when transmitting data samples to external services. In summary, a systematic comparison across five major models demonstrated that well-designed prompts can yield preprocessing code whose outputs largely converge with manually written reference scripts, making the implementation of preprocessing more accessible to a broader range of researchers. Although the present demonstration focused on fundamental preprocessing tasks such as parsing and restructuring process data, the same prompt-based workflow could be extended to more advanced preprocessing operations, including feature engineering, sequence pattern extraction, and the construction of process indicators. These extensions, however, would require separate validation, as they involve more interpretive and analytically consequential transformations.

\subsection{Preprocessed Data Structure}\label{subsec:preprocess_data}

\begin{figure}
    \centering
    \includegraphics[width=1\linewidth]{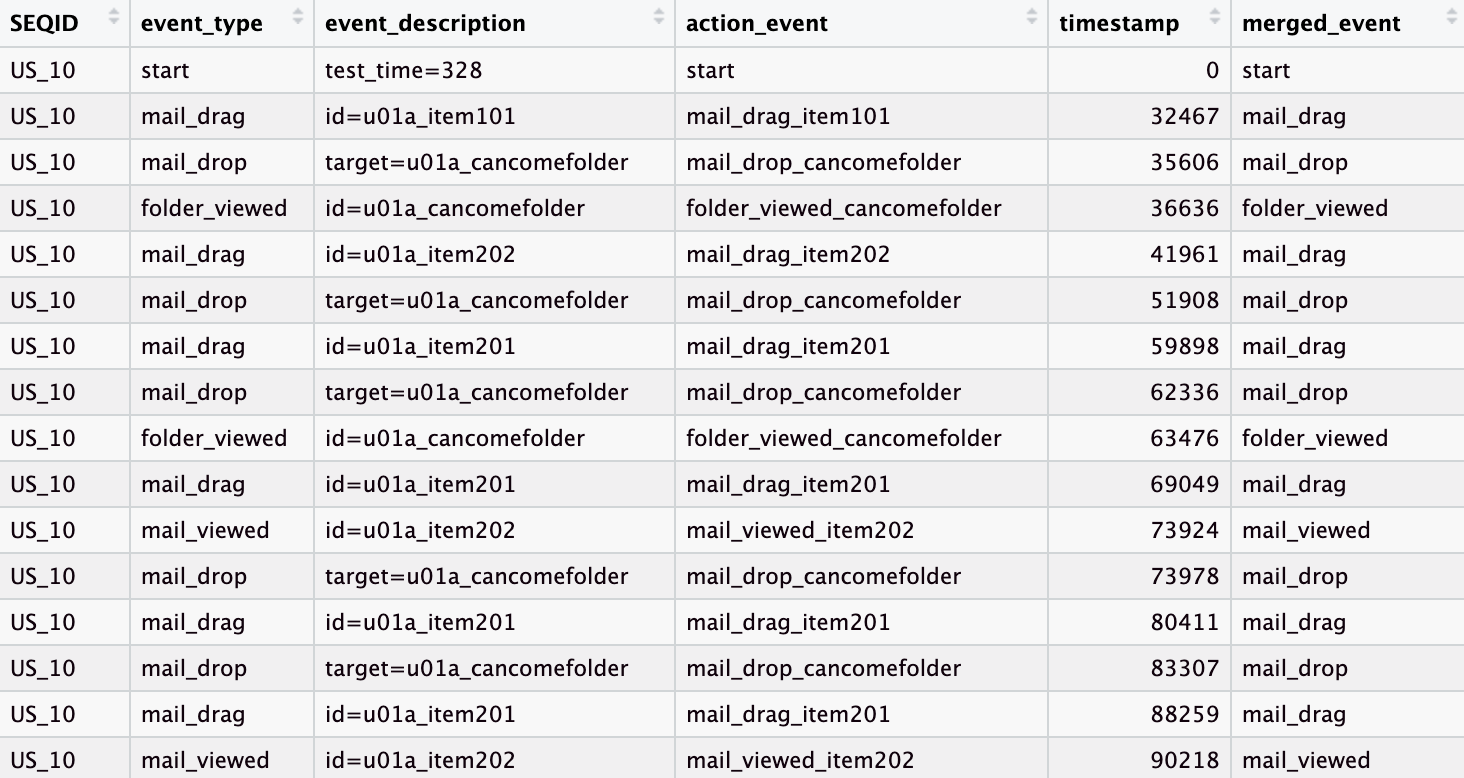}
    \caption{
    Excerpt of preprocessed event-level data for the item \textit{Party Invitations -- Can/Cannot Come} (U01a), United States sample. Compared with the raw log in Figure~\ref{fig:rawlog}, timestamp reversals have been corrected, duplicate records removed, and ancillary actions consolidated with core actions. The \texttt{action\_event} column provides a standardized label combining \texttt{event\_type} and \texttt{event\_description}; the \texttt{merged\_event} column further strips item-specific suffixes to yield coarser action categories used in model-based analyses.
    }
    \label{fig:preprocessed}
\end{figure}

Following the preprocessing steps described above, the raw event-level logs are transformed into a structured, analysis-ready format. Figure~\ref{fig:preprocessed} shows an excerpt of the preprocessed data for the U.S. sample of item \textit{Party Invitations -- Can/Cannot Come} (U01a), which serves as the running example throughout this paper.

Compared with the raw log shown in Figure~\ref{fig:rawlog}, the preprocessed data exhibit several key differences. Timestamp reversals have been corrected, and events are ordered strictly chronologically within each \path{SEQID}. Duplicate records have been removed, and ancillary actions have been consolidated with their corresponding core actions into single action blocks. The \path{action_event} column combines \path{event_type} and the analytically relevant portion of \path{event_description} into a standardized action label that preserves item-specific detail (e.g., \path{mail_drag_item101}). The resulting data frame contains one row per action, with columns \path{SEQID}, \path{event_type}, \path{event_description}, \path{action_event}, and \path{timestamp}, and serves as the common input for the feature-based methods introduced in Section~\ref{sec:method_feature}.

Depending on the analytical goal, researchers may further consolidate \path{action_event} labels into coarser behavioral categories. In this tutorial, model-based methods (Section~\ref{sec:method_advance}) use a \path{merged_event} column derived by stripping item-specific suffixes (e.g., \path{mail_viewed_item101}, \path{mail_viewed_item202} $\to$ \path{mail_viewed}; \path{button_endtask_txt3} $\to$ \path{button_end_confirm}, \path{button_endtask_txt4} $\to$ \path{button_end_cancel}), which reduces the action space and yields more interpretable latent state structures. This recoding is one possible consolidation scheme; researchers should adapt it to the granularity required by their specific analytical context. The recoding script used in this tutorial is available at \url{https://github.com/hwangdabb/log-tutorial/blob/main/S6_Preprocess.R}.

\section{Notation and Preliminary Definitions}\label{sec:notation}

To facilitate a unified presentation across the analytical methods introduced in Sections~\ref{sec:method_feature} and~\ref{sec:method_advance}, we establish a common notational framework here. Method-specific quantities are introduced locally as needed.

We index respondents by $i = 1, \dots, N$ and assessment items by $j = 1, \dots, J$. Each item is associated with a finite set of possible action types $\mathcal{A} = \{1, \dots, M\}$, where $M$ denotes the total number of distinct action categories for that item. Action types are defined after preprocessing and represent meaningful behavioral units such as clicks, drags, or keystrokes. For feature-based methods, $\mathcal{A}$ is defined over \path{action_event} labels; for model-based methods, $\mathcal{A}$ is defined over the consolidated \path{merged_event} labels. The process data for respondent $i$ on item $j$ are represented as a finite sequence of discrete actions,
\[
\mathbf{s}_{ij} = (a_1, a_2, \dots, a_{T_{ij}}),
\]
where $a_t \in \mathcal{A}$ denotes the action taken at step $t$ and $T_{ij}$ denotes the total number of recorded meaningful actions. When the respondent and item indices are clear from context, we suppress the subscripts and write $\mathbf{s} = (a_1, \dots, a_T)$. The prefix subsequence up to step $t$ is denoted $a_{1:t} = (a_1, \dots, a_t)$.

Each action is associated with a timestamp, and the complete timestamped interaction record for respondent $i$ on item $j$ is represented as
\[
0 = t_{ij}^{(0)} < t_{ij}^{(1)} < \cdots < t_{ij}^{(T_{ij})} < t_{ij}^{(\text{sub})},
\]
where $t_{ij}^{(0)}$ is the item presentation time, $t_{ij}^{(k)}$ is the time at which action $a_k \in \mathbf{s}_{ij}$ is recorded for $k = 1, \dots, T_{ij}$, and $t_{ij}^{(\text{sub})}$ is the submission time.

\section{Feature-based Methods and Applications}\label{sec:method_feature}

The analysis of process data typically begins with transforming raw action sequences into interpretable representations that can be used for subsequent modeling and inference. This section reviews fundamental approaches to feature extraction, progressing from simple aggregate indicators to more sophisticated methods that preserve sequential structure and reduce dimensionality. Each approach offers a different trade-off between interpretability, computational complexity, and the richness of behavioral information captured. For each method, this section introduces the methodological background and underlying rationale. Empirical illustrations using the U.S. PIAAC sample, together with reproducible \path{R} implementations, are provided in the Supplementary Materials. For descriptive process indicators, both the application and implementation are presented in Section~S2 of the Supplementary Materials, whereas the remaining methods are illustrated in Section~S3--S7 of the Supplementary Materials.

\subsection{Descriptive Process Indicators}\label{sec:basic_indicators}

Basic process indicators provide aggregate summaries that capture fundamental behavioral patterns in process data without explicitly modeling the sequential structure of actions. These indicators serve as a natural starting point for exploratory analysis, allowing researchers to identify broad differences across respondents or items before applying more complex sequence-based or model-based techniques.  Despite their simplicity, basic process indicators often yield substantively meaningful insights when interpreted with appropriate contextual information.

\subsubsection{Methodological Background}

Time-based indicators summarize how respondents allocate time while interacting with an assessment item, capturing temporal aspects of problem-solving behavior such as cognitive load, engagement, and decision-making processes.

\paragraph*{Time on Task}
Time on task (ToT) measures the total duration a respondent spends on an item and is defined as
\[
\text{ToT}_{ij} = t_{ij}^{(\text{sub})} - t_{ij}^{(0)}.
\]
This fundamental indicator is closely related to the speed--accuracy tradeoff and has been widely used as a proxy for cognitive effort, problem difficulty, or engagement level \citep{goldhammertime, kupiainen2014role, goldhammer2017relating}. Context is therefore crucial: a long time on task accompanied by many actions suggests active exploration, whereas a long duration with few actions may indicate disengagement, off-task behavior, or external interruptions.

\paragraph*{Time to First Action}
Time to first action (TFA) captures the latency between item presentation and the first meaningful action,
\[
\text{TFA}_{ij} = t_{ij}^{(1)} - t_{ij}^{(0)}.
\]
This indicator reflects initial comprehension speed and the time required to formulate an initial problem-solving strategy. Extremely short latencies may suggest impulsive behavior or insufficient planning, whereas unusually long latencies may indicate difficulty in understanding task instructions or interface demands \citep{kaldes2024s}. If no meaningful action is observed ($T_{ij} = 0$), $\text{TFA}_{ij}$ is undefined; in practice, such cases may be treated as missing or handled separately as rapid submissions depending on the research question.

\paragraph*{Number of Actions}
Number of actions (NoA) counts the total number of meaningful interactions performed by respondent $i$ on item $j$ during problem solving, formally defined as
\[
\text{NoA}_{ij} = T_{ij}.
\]
This indicator captures overall activity level and provides a coarse summary of behavioral complexity. The relationship between action counts and performance is often nonlinear: very few actions may indicate insufficient engagement or rapid guessing, while excessive actions may reflect inefficient trial-and-error strategies \citep{teig2020identifying, he2021leveraging}.

Overall, such indicators should rarely be interpreted in isolation. Their substantive meaning becomes clearer when analyzed jointly with action-based or sequence-based features, which we discuss in the subsequent sections.

\subsubsection{Application to PIAAC Process Data}

Basic process indicators have been widely applied in PIAAC research to characterize respondent behavior and link it to proficiency. Using PIAAC data, \citet{goldhammertime, goldhammer2017relating} demonstrated that the relationship between ToT and accuracy varies systematically across domains: in reading tasks requiring more automatic processing, faster responses tend to be more accurate, while in problem-solving tasks requiring controlled processing, taking more time is associated with higher accuracy. This relationship is further moderated by task difficulty and respondent ability \citep{goldhammertime, goldhammer2017relating}: for difficult items and low-ability respondents, investing more time has a positive effect on response accuracy, whereas for easy items and high-ability respondents, the effect becomes negative \citep{kramer2023testing}. 

For NoA, \citet{goldhammer2017relating} demonstrated that the relationship between action counts and task success follows an inverted-U pattern: taking few actions corresponds to passive disengagement, a moderate number reflects active engagement, and excessively high counts suggest distraction or disorientation within the task environment. This pattern is further moderated by task demands---in tasks requiring more steps to complete, a strong positive association emerged between action counts and success probability, whereas in tasks requiring short navigation paths, the association was considerably weaker. Together, these findings suggest that NoA should be interpreted with reference to both the range of values and the navigational demands of the task at hand.

While each basic indicator provides standalone information, its substantive value emerges most clearly through joint interpretation. For example, a long response time paired with few actions suggests different behavior than a long response time with many actions: the former may indicate careful planning or hesitation, whereas the latter reflects active exploration. This illustrates why relying on a single indicator can be misleading in process data analysis. Taken together, basic process indicators establish a foundation for understanding response behaviors while deliberately abstracting away from sequential structure. The following sections introduce methods that leverage the ordering of actions, beginning with n-gram features, which summarize local sequential patterns in a simple and flexible way.

Empirical illustrations and reproducible \path{R} implementation are provided in Section~S2 of the Supplementary Materials, with full code available at \url{https://github.com/hwangdabb/log-tutorial/blob/main/S5_indicator.R}.

\subsection{N-gram Analysis}\label{subsec:ngram}

N-gram analysis extends aggregate indicators by decomposing each action sequence into overlapping contiguous subsequences, thereby preserving local ordering information. Originating from computational linguistics \citep{brown1992class, cavnar1994n}, n-gram features provide a simple and flexible way to quantify local behavioral regularities that may differentiate strategies or predict outcomes \citep{he2015identifying, he2016analyzing}.

Following the notation introduced in Section~\ref{sec:notation}, let $\mathbf{s}_{ij} = (a_1, \ldots, a_{T_{ij}})$ denote the action sequence for respondent $i$ on item $j$. For a fixed order $n\ge 1$, the set of overlapping n-grams extracted by a sliding window is
\[
n\text{-grams}(\mathbf{s}_{ij})
= \bigl\{\,(a_t, a_{t+1}, \ldots, a_{t+n-1})
\;:\; t = 1, \ldots, T_{ij} - n + 1\,\bigr\}.
\]
Unigrams ($n=1$) reduce to action frequencies, bigrams ($n=2$) capture immediate transitions, and larger $n$ encode longer local patterns but yield increasingly sparse representations as $n$ grows. In practice, sequences are typically augmented with special boundary tokens such as \path{START} and \path{END} to preserve initial and terminal behaviors \citep{zhou2024investigating}. Boundary tokens enable the capture of entry and exit behaviors that would otherwise be undetectable: for example, a \path{START}–action bigram identifies respondents who begin with a particular action as their very first step, a pattern that standard n-gram extraction from sequence interiors cannot recover.

Not all extracted patterns contribute equally to distinguishing respondents or predicting outcomes. Let $\mathcal{V}_n$ denote the set of distinct n-grams observed across all $N$ sequences, and index a pattern by $v \in \mathcal{V}_n$. Each sequence can then be encoded as a sparse feature vector over $\mathcal{V}_n$ using the within-sequence frequency $tf_{v,ij}$ (i.e., the number of occurrences of pattern $v$ in $\mathbf{s}_{ij}$) or a binary presence indicator $\mathbb{I}(tf_{v,ij} \ge 1)$.

To improve discriminative power, \citet{he2015identifying, he2016analyzing} adapted inverse document frequency (IDF) to process data and termed it inverse sequence frequency (ISF). This weighting scheme downweights ubiquitous patterns while upweighting rare but potentially informative ones. Specifically,
\[
\mathrm{ISF}_v=\log\!\left(\frac{N}{sf_v}\right)\ge 0,
\]
where $sf_v$ is the number of sequences containing pattern $v$ at least once. Patterns appearing in all sequences have $\mathrm{ISF}_v=0$, while rare patterns receive higher weights. The term frequency--inverse sequence frequency (TF--ISF) weight is then
\[
\mathrm{TF\text{-}ISF}(v, ij) =
\begin{cases}
\left[1 + \log(tf_{v,ij})\right]
\cdot \log\!\left({N}/{sf_v}\right), 
& tf_{v,ij} \ge 1, \\
0, & tf_{v,ij} = 0.
\end{cases}
\]
The logarithmic transformation $1+\log(tf_{v,ij})$ moderates the influence of repeated occurrences within a sequence, while the ISF component highlights patterns that occur selectively across respondents \citep{he2015identifying, he2016analyzing}.

Because the number of candidate patterns can be large even for small $n$, sparsity control is typically required. A standard pipeline proceeds as follows: (i) remove patterns with very small support (e.g., $sf_v < \epsilon$), (ii) discard non-informative patterns with $\mathrm{ISF}_v=0$, and (iii) screen the remaining patterns using a simple association statistic with a target variable \citep{he2015identifying, he2016analyzing}. For a binary outcome (e.g., correct vs.\ incorrect), the Pearson chi-square statistic computed from the $2\times 2$ contingency table of pattern presence versus outcome \citep{yang1997comparative} is
\[
\chi^2_v=\sum_{r'=1}^{2}\sum_{c'=1}^{2}\frac{(O_{r'c'}-E_{r'c'})^2}{E_{r'c'}},
\]
where $O_{r'c'}$ and $E_{r'c'}$ are the observed and expected counts under independence. Retaining patterns with large $\chi^2_v$ yields interpretable candidates for downstream predictive modeling or group comparison, highlighting local action subsequences that are strongly associated with success or failure in problem solving. 
To illustrate the effect of this pipeline on U01a, the candidate feature space was reduced as follows: for unigrams ($n=1$), from 130 to 76 patterns; for bigrams ($n=2$), from 1{,}497 to 497 patterns; and for trigrams ($n=3$), from 5{,}253 to 917 patterns, demonstrating how sparsity grows rapidly with $n$ and how the filtering steps substantially reduce dimensionality before downstream analysis.
Empirical illustrations and reproducible \path{R} implementation are provided in Section~S3 of the Supplementary Materials, with full code available at \url{https://github.com/hwangdabb/log-tutorial/blob/main/S5_ngram.R}.

\subsection{Multidimensional Scaling (MDS)}\label{subsec:mds}

Despite the advantages discussed in the previous section, $n$-grams capture only local sequential structure within a window of size $n$. Consequently, they may miss higher-level strategies spanning non-adjacent actions and often yield high-dimensional, sparse representations. To overcome these limitations, we introduce distance-based representations and multidimensional scaling (MDS). These approaches operate directly on entire action sequences and embed them into a low-dimensional space that preserves pairwise similarities.

Multidimensional scaling (MDS) is a classical dimension-reduction technique that takes a pairwise dissimilarity matrix as input and produces low-dimensional coordinates. These coordinates approximately preserve between-sequence dissimilarities, enabling both visualization and downstream modeling \citep{kruskal1978multidimensional, davison2000multidimensional}. The goal of MDS is to locate objects in a vector space according to their pairwise dissimilarities such that similar objects are placed close together, while dissimilar objects are placed farther apart.

The key to applying MDS to process data is the choice of a dissimilarity measure between action sequences. Once a dissimilarity $d(\cdot,\cdot)$ is specified, MDS yields an embedding in which sequences similar under $d$ are located nearby. With an appropriate rotation, each extracted feature can be interpreted as describing the variation of a certain ability or behavioral pattern among respondents \citep{tang2020latent}, although substantive interpretations should be supported by associations with external variables such as item outcomes or proficiency measures.

Since an appropriate dissimilarity measure for action sequences should accommodate (i) discreteness of action types that do not admit arithmetic operations, (ii) variable sequence lengths, and (iii) sensitivity to action order, \citet{tang2020latent} adopt the following dissimilarity measure, originally proposed by \citet{gomez2008similarity}. Consider two respondents $i$ and $i'$ on a fixed item $j$, with action sequences $\mathbf{s}_{ij} = (a_1, \ldots, a_{T_{ij}})$ and $\mathbf{s}_{i'j} = (a_1, \ldots, a_{T_{i'j}})$ of lengths $T_{ij}$ and $T_{i'j}$. The dissimilarity between $\mathbf{s}_{ij}$ and $\mathbf{s}_{i'j}$ is defined as
\begin{equation}\label{eq:d}
    d(\mathbf{s}_{ij}, \mathbf{s}_{i'j})
    = \frac{f(\mathbf{s}_{ij}, \mathbf{s}_{i'j}) + 
    g(\mathbf{s}_{ij}, \mathbf{s}_{i'j})}
    {T_{ij} + T_{i'j}},
\end{equation}
where $f(\mathbf{s}_{ij}, \mathbf{s}_{i'j})$ quantifies the dissimilarity among actions that appear in both sequences, and $g(\mathbf{s}_{ij}, \mathbf{s}_{i'j})$ counts actions appearing in only one sequence. For each action type $a \in \mathcal{A}$, let $\mathbf{s}_{ij}^{\,a}$ denote the ordered sequence of positions at which $a$ occurs in $\mathbf{s}_{ij}$, with length $T_{ij}^a$. Partitioning actions into those common to both sequences ($C_{ii'j}$) and those unique to each ($U_{ii'j}$ and $U_{i'ij}$), $f$ and $g$ are defined as
\begin{equation}\label{eq:g}
    f(\mathbf{s}_{ij}, \mathbf{s}_{i'j}) = \frac{\displaystyle\sum_{a \in C_{ii'j}}  \sum_{k=1}^{K_{ii'j}^{a}} \bigl|\mathbf{s}_{ij}^{\,a}(k) - \mathbf{s}_{i'j}^{\,a}(k)\bigr|}{\max\{T_{ij}, T_{i'j}\}}, \quad g(\mathbf{s}_{ij}, \mathbf{s}_{i'j}) = \sum_{a \in U_{ii'j}} T_{ij}^a + \sum_{a \in U_{i'ij}} T_{i'j}^a,
\end{equation}
where $K_{ii'j}^{a} = \min\{T_{ij}^a, T_{i'j}^a\}$ and $\mathbf{s}_{ij}^a(k)$ denotes the position of the $k$-th occurrence of $a$ in $\mathbf{s}_{ij}$.

Intuitively, $f$ measures how the serial positions of common actions differ between the two sequences, while $g$ penalizes actions that appear in only one sequence. Other dissimilarity measures are also possible (e.g., edit-distance or alignment-based measures), and the resulting MDS configuration should be interpreted as conditional on the selected measure \citep{tang2020latent}.

For a fixed item $j$, let $\mathbf{D}_j = (d_{ii'j})$ denote the $N_j \times N_j$ pairwise dissimilarity matrix, where $N_j$ is the number of respondents who answered item $j$. MDS maps each sequence $\mathbf{s}_{ij}$ to a latent vector $\boldsymbol{\phi}_{ij} \in \mathbb{R}^K$ by minimizing the least-squares criterion
\[
\sum_{i < i'} \bigl(d_{ii'j} - 
\|\boldsymbol{\phi}_{ij} - 
\boldsymbol{\phi}_{i'j}\|\bigr)^2
\]
with respect to $\boldsymbol{\Phi}_j = (\boldsymbol{\phi}_{1j}, \ldots, \boldsymbol{\phi}_{N_j j})^T$. The feature extraction procedure \citep{tang2020latent} proceeds as follows: (1) compute $\mathbf{D}_j$ for all $N_j(N_j-1)/2$ sequence pairs on item $j$; (2) obtain $K$ raw MDS features by minimizing the criterion above via stochastic approximation \citep{robbins1951stochastic}; and (3) apply PCA to yield $K$ principal features $\boldsymbol{\Phi}_j$. The number of features $K$ can be selected via $m$-fold cross-validation on the pairwise dissimilarities. For the illustrative item U01a, we follow \citet{tang2020latent}, who selected $K = 50$ via fivefold cross-validation on the same dataset. Empirical illustrations and reproducible \path{R} implementation are provided in Section~S4 of the Supplementary Materials, with full code available at \url{https://github.com/hwangdabb/log-tutorial/blob/main/S5_MDS.R}.

\subsection{DIF Analysis}\label{subsec:dif}

Differential item functioning (DIF) occurs when respondents from different groups who are matched on the target ability have unequal probabilities of answering an item correctly \citep{zumbo1999handbook, penfield2000assessing}, thereby undermining the fairness of test-based inferences \citep{camilli2018ongoing}. Following \citet{qi2023process}, consider a general item response function (IRF) for respondent $i$ on item $j$,
\begin{equation*}
    p_j(\theta_i, \boldsymbol{\eta}_i) \triangleq P(Y_{ij} = 1 \mid \theta_i, \boldsymbol{\eta}_i),
\end{equation*}
where $\theta_i$ denotes the target latent trait---in the PS-TRE context, problem-solving proficiency---and $\boldsymbol{\eta}_i$ is a nuisance attribute vector capturing construct-irrelevant characteristics. DIF arises when the conditional distribution of $\boldsymbol{\eta}_i$ given $\theta_i$ differs across groups $g \in \{r, f\}$, so that the marginal IRF with $\boldsymbol{\eta}$ integrated out satisfies $p_j(\theta_i, g{=}r) \neq p_j(\theta_i, g{=}f)$. In the PS-TRE context, for instance, $\boldsymbol{\eta}_i$ captures interface-operation preferences such as drag-and-drop usage: older respondents tend to rely on this strategy less frequently than younger ones, so the grouping variable (age) creates a performance gap through its correlation with $\boldsymbol{\eta}_i$, not through any difference in problem-solving competency itself.

Two classical approaches identify items exhibiting DIF. The Mantel--Haenszel (MH) procedure \citep{wainer1988testvalidity} stratifies respondents into $k$ score intervals and compares item performance across groups via the common odds ratio
\begin{equation*}
    \alpha_{MH} = \frac{\sum_k (C_{rk} I_{fk} / N_k)}{\sum_k (I_{rk} C_{fk} / N_k)},
\end{equation*}
where $C_{gk}$ and $I_{gk}$ denote correct and incorrect response counts for group $g$ at interval $k$, and $N_k$ is the stratum total; the null hypothesis is $\alpha_{MH} = 1$. The IRT-based likelihood-ratio test compares a constrained model $M_c$, which fixes item parameters across groups, against an augmented model $M_v$ that allows them to vary, via $G^2 = 2\ln[L_v/L_c]$, which follows an approximate chi-square distribution under the null. While both approaches reliably flag DIF items, neither explains which behavioral mechanisms drive the disparity.

Process data bridge this gap by providing observable proxies for the latent $\boldsymbol{\eta}_i$. MDS-derived behavioral coordinates introduced in Section~\ref{subsec:mds}, which compactly encode construct-irrelevant tendencies that may vary systematically across demographic groups, serve as such proxies. Let $\mathbf{X}_{ij} = (X_{ij1}, \ldots, X_{ijK})^\top$ denote the $K$-dimensional MDS feature vector for respondent $i$ on item $j$. \citet{qi2023process} substitute these for the latent nuisance trait, yielding an augmented IRF
\begin{equation*}
    P(Y_{ij} = 1 \mid \theta_i, \mathbf{X}_{ij})
    = \sigma\!\left(\delta_j\,\theta_i + \sum_{k=1}^{K} \gamma_{jk}\,X_{ijk} + b_j\right),
\end{equation*}
where $\sigma(\cdot)$ is the logistic function, $\delta_j$ is the discrimination parameter, $b_j$ is the intercept, and $\gamma_{jk}$ are the loadings on the process features. If the included features adequately capture $\boldsymbol{\eta}_i$, this IRF becomes identical across groups. The residual group disparity is quantified by the empirical $L_2$ distance between the estimated IRFs of the focal and reference groups,
\begin{equation*}
    \hat{d}_j^2 = \sum_{i=1}^N \left[\hat{p}_{fj}(\theta_i, \mathbf{X}_{ij})
                - \hat{p}_{rj}(\theta_i, \mathbf{X}_{ij})\right]^2,
\end{equation*}
where $\hat{p}_{fj}$ and $\hat{p}_{rj}$ denote the fitted IRFs for the focal and reference groups, respectively; the asymptotic null distribution of $\hat{d}_j^2$ follows a generalized chi-square distribution. A forward stepwise procedure then incrementally adds one feature at a time---selecting the candidate that most reduces $\hat{d}_j$, subject to coefficient significance and estimability of $\theta$---until $\hat{d}_j$ falls below the significance threshold. The DIF-corrected target trait is then estimated via maximum likelihood over all items,
\begin{equation*}
    \hat{\theta}_i^* = \operatorname*{arg\,max}_{\theta}
    \sum_{j=1}^J \log P(Y_{ij} = y_{ij} \mid \theta, \mathbf{X}_{ij}),
\end{equation*}
where DIF-flagged items use the augmented IRF with the selected features, and anchor items (i.e., non-DIF items used to anchor the ability scale) retain the standard 2PL IRF.

\citet{chen2025reducing} propose a complementary GLM-based approach that accommodates multiple demographic groupings simultaneously. Rather than selecting features sequentially, a nuisance surrogate $\hat{\eta}_i = \mathbf{w}^\top \mathbf{X}_i$ is constructed from all available MDS features $\mathbf{X}_i = (X_{i1}, \ldots, X_{iK})^\top$. The weight vector $\mathbf{w}$ is estimated by minimizing the log-likelihood difference between a grouping-invariant model conditioning on $(\hat{\theta}_i, \hat{\eta}_i)$ and a group-varying model that additionally includes a group main effect, so that the resulting surrogate, once conditioned on, renders the grouping variable redundant. In the linear case, a closed-form solution exists; for nonlinear models or multiple grouping variables, numerical optimization is used. A new scoring rule is then derived that combines $\hat{\theta}_i$ and $\hat{\eta}_i$ such that the final score reflects only the primary dimension, thereby correcting for construct-irrelevant behavioral differences at the individual level. Empirical illustrations and reproducible \path{R} implementation are provided in Section~S5 of the Supplementary Materials, with full code available at \url{https://github.com/hwangdabb/log-tutorial/blob/main/S5_DIF.R}.

\section{Model-based Methods}\label{sec:method_advance}

While the feature-based methods in Section~\ref{sec:method_feature} provide flexible and computationally efficient representations of behavioral patterns, they do not explicitly model the latent cognitive processes that generate observable actions. The model-based methods introduced in this section address this limitation by positing explicit probabilistic structures for how problem-solving states evolve over time.

These methods also operate on a different level of action representation. Whereas the feature-based analyses use fine-grained \path{action_event} labels that retain item-specific detail, the model-based analyses aim to identify latent behavioral stages shared across respondents rather than object-specific interaction patterns. Retaining that level of specificity would inflate the action space, produce sparse emission distributions in HMMs, and yield less interpretable action frequency profiles in SIP. Accordingly, the analyses in Sections~\ref{subsec:hmm} and~\ref{subsec:sip} use the consolidated \path{merged_event} labels introduced in Section~\ref{subsec:preprocess_data}.

\subsection{Hidden Markov Models (HMMs)}\label{subsec:hmm}

Hidden Markov models (HMMs) are a latent-state framework well suited to analyzing process data, where an observed action sequence is treated as an observable trace of an unobserved sequence of cognitive or behavioral states \citep{arieli2019understanding, tang2024latent, rodriguez2015discovering}. Unlike feature-based methods that summarize action sequences into fixed-dimensional representations without explicitly modeling the underlying cognitive process, HMMs posit an explicit probabilistic structure for how latent problem-solving states evolve over time to generate observable actions, yielding interpretable parameters that describe (i) which actions are characteristic of each latent stage and (ii) how respondents transition among stages over the course of problem solving \citep{xiao2021exploring, liu2025uncovering}.

Formally, let $\mathbf{s}_{ij} = (a_1, \ldots, a_{T_{ij}})$ denote the observed action sequence for respondent $i$ on item $j$ as defined in Section~\ref{sec:notation}. The corresponding latent state sequence is $S_{1:T_{ij}} = (S_1, \ldots, S_{T_{ij}})$ with $S_t \in \{1, \ldots, Q\}$ for a prespecified number of latent states $Q$. An HMM assumes a first-order Markov property for the states and conditional independence of actions given states \citep{cappe2005inference, eddy1996hidden, eddy2004hidden}, so that the joint distribution can be written as
\begin{align}
    P(a_{1:T_{ij}},\, S_{1:T_{ij}})
    =
    P(S_1)\,P(a_1 \mid S_1)
    \prod_{t=2}^{T_{ij}} P(S_t \mid S_{t-1})\,P(a_t \mid S_t),
\end{align}
where $P(S_1)$ is the initial-state distribution, $P(S_t \mid S_{t-1})$ is the transition probability matrix governing how states evolve over time, and $P(a_t \mid S_t)$ is the emission probability matrix specifying which actions are likely within each latent state. Parameter estimation is typically performed via maximum likelihood using the Baum--Welch algorithm, an expectation--maximization (EM) procedure that employs forward--backward recursions to iteratively update parameter estimates until convergence \citep{baum1970maximization, baggenstoss2001modified}. Given a fitted model, the most probable latent-state sequence underlying each observed action sequence can be recovered using the Viterbi algorithm \citep{viterbi2003error, churbanov2008implementing}, or posterior state probabilities may be used when uncertainty in state assignment is important \citep{luong2013fast}.

The number of latent states $Q$ is selected by fitting models across a prespecified range and comparing AIC and BIC, balancing statistical fit against interpretability \citep{visser2002fitting, dridi2018akaike, chien2005predictive}. 
For U01a, both criteria decreased monotonically across $Q=1$ to $9$; we accordingly report results for $Q = 9$ as the upper boundary of the candidate range examined, though 
a smaller $Q$ may be preferred on interpretability grounds (see Figure~S6 in the Supplementary Materials).

State interpretation rests primarily on the emission probability matrix, which quantifies the likelihood of each action being produced by each latent state. In process data analysis, states are labeled by examining which actions carry the highest emission probabilities, and substantive meaning is inferred from the characteristic behavioral pattern---for example, a state dominated by sorting and searching actions might be labeled as ``use of tools'' \citep{xiao2021exploring}. Because HMM state labels are identifiable only up to permutation---that is, relabeling states does not change model fit---comparisons across groups or items should rely on substantively defined emission profiles rather than arbitrary numeric labels. The transition matrix complements the emission matrix by revealing how respondents move through stages over time: a high probability of remaining in the same state suggests that respondents tend to repeat a particular action multiple times before moving on (e.g., applying a sorting function repeatedly), whereas frequent transitions to other states may reflect checking behavior, backtracking to earlier stages, or uncertainty about the next step. In more advanced formulations, the HMM framework can be augmented with covariates---such as goal orientation or partial solution state---to model individual differences in initial-state and transition probabilities \citep{arieli2019understanding}. Empirical illustrations and reproducible \path{R} implementation are provided in Section~S6 of the Supplementary Materials, with full code available at \url{https://github.com/hwangdabb/log-tutorial/blob/main/S6_HMM.R}.

\subsection{Subtask Identification Procedure (SIP)}\label{subsec:sip}

Process data generated in technology-rich environments often consist of lengthy and highly variable action sequences, making it difficult to apply standard statistical tools directly.
A natural approach to managing this complexity is to decompose each sequence into shorter, coherent segments before analysis---a strategy with a long tradition in sequence modeling and natural language processing \citep{hearst1994multi}. The subtask identification procedure \citep[SIP;][]{wang2023subtask} adapts this idea to process data by segmenting a long action sequence into a series of shorter, meaningful subsequences---referred to as \textit{subtasks}---thereby reducing dimensionality while preserving the sequential structure of problem-solving behavior. The procedure rests on a central observation: actions \textit{within} a subtask tend to be highly predictable, since respondents are executing a coherent local procedure, whereas transitions \textit{between} subtasks are associated with elevated uncertainty, as the respondent may shift focus to any of several possible next steps. SIP operationalizes this intuition through three sequential steps: action prediction, entropy-based segmentation, and subtask clustering.

The first step fits a predictive model for the next action given the respondent's action history.
Let $\mathbf{s} = (a_1, \ldots, a_T)$ denote the observed action sequence, where $a_t \in \mathcal{A}$ is the action at step $t$ from a discrete action space $\mathcal{A}$ with $M$ possible actions (as defined in Section~\ref{sec:notation}).
The action history $a_{1:t}$ is first compressed into a $K$-dimensional latent vector $\boldsymbol{u}_t$ by a recurrent neural network \citep[RNN;][]{elman1990finding}---specifically a gated recurrent unit \citep[GRU;][]{cho2014learning}, which captures sequential dependencies without an unbounded parameter expansion.
The GRU was adopted following the original SIP formulation of \citet{wang2023subtask}.
Compared with LSTMs, GRUs have fewer parameters and are computationally more efficient, making them well suited for the relatively short action sequences typical of PS-TRE items.
Compared with attention-based Transformers, which process the entire sequence at once, GRUs read actions one step at a time in order, which naturally reflects how problem-solving behavior unfolds sequentially.
The conditional probability of the next action is then given by a multinomial logistic model,
\begin{equation}
    p_{tm} = P(a_{t+1}=m \mid a_{1:t}) = \begin{cases}
        \frac{\exp(\boldsymbol{\beta}_m^\top \boldsymbol{u}_t + \alpha_m)}{1 + \sum_{\ell=1}^{M-1} \exp(\boldsymbol{\beta}_\ell^\top \boldsymbol{u}_t + \alpha_\ell)}, & \quad m=1, \dots, M-1
        \\ 
        \frac{1}{1 + \sum_{\ell=1}^{M-1} \exp(\boldsymbol{\beta}_\ell^\top \boldsymbol{u}_t + \alpha_\ell)}, & \quad m=M
    \end{cases}
\end{equation}
where $\{\alpha_m, \boldsymbol{\beta}_m\}$ are learnable parameters estimated by maximizing the log-likelihood of observed sequences.

Once the predictive model is fitted, the second step quantifies the uncertainty of its predictions at each time point via the Shannon entropy \citep{bromiley2004shannon},
\begin{equation}
  H_t \;=\; -\sum_{m=1}^{M} p_{tm} \log p_{tm}.
\end{equation}
The resulting entropy sequence $\mathbf{h} = (H_1, \ldots, H_{T-1})$ typically traces a series of U-shaped curves: entropy is low within a subtask and rises at transitions between subtasks.
Subtask boundaries are identified at locations where the entropy sequence exhibits a sufficiently deep local peak, formally defined through the concept of a \textit{U-curve}. A subsequence $H_{\ell:r}$ of $\mathbf{h}$ is said to form a U-curve if
\begin{equation}
  \min\{H_\ell,\, H_r\} \;-\; \min_{\ell \le t \le r} H_t
  \;\ge\;
  \lambda\,\Bigl(\max_t H_t - \min_t H_t\Bigr),
\end{equation}
where $\lambda \in [0,1]$ controls the minimum relative depth. At the extremes, $\lambda = 0$ qualifies every subprocess between two consecutive local maxima as a U-curve, yielding maximally fine-grained segmentation, whereas $\lambda = 1$ and the global maximum of $\mathbf{h}$ is unique, no subsequence qualifies as a U-curve, treating the entire sequence as a single subtask. In practice, the choice of $\lambda$ should be guided by the expected granularity of subtask structure, and sensitivity to this parameter warrants examination in applied settings. Segmentation points are then selected from the local maxima of $\mathbf{h}$ via a bidirectional filtering algorithm, which prevents extended high-entropy intervals between subtasks from being mistakenly fragmented \citep{wang2023subtask}.

The third step clusters the segmented subprocesses into a prespecified number of subtask types. Each subprocess $k$ is summarized by an \textit{action frequency profile} $\mathbf{z}_k = (z_{k 1}, \dots, z_{k M})^\top$, whose $m$-th entry records the relative frequency of action $m$ within that subprocess. Here, $\mathbf{z}_k$ is a probability vector with $0 \leq z_{k m} \leq 1, \; m=1, \dots, M$ and $\sum_{m=1}^M z_{k m} = 1$. Then, all profiles are grouped via k-means with a prespecified number of clusters $R_c$ and the Hellinger distance $d_H(\mathbf{p},\mathbf{q}) = [\sum_m(\sqrt{p_m}-\sqrt{q_m}\,)^2]^{1/2}$, where $R_c$ is determined in advance based on domain knowledge or prior theoretical expectations about the task structure. Each cluster is then assigned a substantive label based on its most frequent actions, condensing the original action sequence into a short \textit{subtask sequence} that is directly amenable to visualization and downstream analysis. Empirical illustrations and reproducible \path{R} implementation are provided in Section~S7 of the Supplementary Materials, with full code available at \url{https://github.com/hwangdabb/log-tutorial/blob/main/S6_SIP.R}.

\section{Discussion}\label{sec:conclusion}
This paper addressed three gaps in the existing process data literature through a unified analytical pipeline covering preprocessing, feature-based methods, and model-based approaches. First, to address the lack of systematic preprocessing evaluation, we presented a structured pipeline and demonstrated through a five-model LLM comparison that well-designed prompts yield preprocessing outputs that largely converge with manually written reference implementations. Second, to address the absence of an integrated analytical view, we showed how preprocessing decisions directly shape downstream action representations and how feature-based and model-based methods operate on complementary levels of behavioral abstraction. Third, to address the scarcity of cross-method consistency checks, we applied multiple analytical approaches to a common dataset and examined whether similar substantive conclusions emerge across different analytical families.

The six methods differ substantially in their assumptions, behavioral information captured, computational demands, and target research questions; Table~\ref{tab:method_comparison} provides a structured comparison. In practice, method selection should be guided by the research question rather than by analytical preference. Descriptive indicators suit exploratory characterization; n-gram analysis with TF--ISF weighting is efficient for identifying local behavioral motifs; MDS yields compact, fixed-dimensional representations compatible with downstream regression or IRT-based correction; HMMs are preferred when the goal is to model latent cognitive stages explicitly; SIP is better suited for task-level segmentation and strategy identification; and process-informed DIF analysis links behavioral observations directly to psychometric fairness. These methods are not mutually exclusive. The present paper illustrates one productive integration in which MDS-derived features served as nuisance proxies in the DIF correction of Section~\ref{subsec:dif}.

\begin{table}[htb]
\centering
\caption{
Comparison of the six analytical methods introduced in this paper. \texttt{action\_event} labels retain item-specific suffixes (e.g., \texttt{mail\_drag\_item101}); \texttt{merged\_event} labels collapse these into coarser categories (e.g., \texttt{mail\_drag}). See Section 3.3 for details.
}
\label{tab:method_comparison}
\scriptsize
\begin{tabular}{p{1.3cm} p{2.4cm} p{2.4cm} p{2.0cm} 
                p{2.0cm} p{1.0cm} p{1.8cm}}
\toprule
Method 
  & Primary Research Question 
  & Key Assumptions 
  & Strengths 
  & Limitations 
  & Action Label
  & Computational Note \\
\midrule
Process Indicators
  & How much time and effort did respondents invest?
  & None beyond timestamped sequences
  & Simple, interpretable, widely comparable
  & Ignores sequential structure; misleading in isolation
  & \path{action_event}
  & --- \\[6pt]
N-gram
  & Which local action patterns discriminate performance groups?
  & Local sequential patterns carry discriminative information
  & Preserves local order; flexible ($n$); directly interpretable
  & Ignores long-range structure; sparse for large $n$
  & \path{action_event}
  & --- \\[6pt]
MDS
  & What compact behavioral representation best predicts outcomes?
  & Dissimilarity measure adequately captures behavioral distance
  & Compact representation; strong predictive validity
  & Interpretation requires external validation; sensitive to dissimilarity choice
  & \path{action_event}
  & $O(N^2)$ pairwise distance; expensive for large $N$ \\[6pt]
DIF
  & Do construct-irrelevant group differences bias item performance?
  & MDS features adequately proxy latent nuisance trait; item responses follow a logistic IRT model
  & Addresses fairness; links behavioral to psychometric analysis
  & Requires predefined grouping variable; stepwise selection can be unstable
  & \path{action_event}
  & --- \\
  [6pt]
HMM
  & What latent cognitive stages underlie observed sequences?
  & First-order Markov property; conditional independence of actions given states
  & Explicit latent state model; interpretable parameters; extensible to covariates
  & $Q$ pre-specified; IC often decreases monotonically; state labeling post-hoc
  & \path{merged_event}
  & Fit over range of $Q$; cost scales with candidates \\[6pt]
SIP
  & What subtask strategies do respondents employ?
  & Within-subtask predictability exceeds between-subtask; GRU captures dependencies
  & Data-driven segmentation; handles variable-length sequences; intuitive output
  & $\lambda$ not principled; hard to evaluate without reference labels
  & \path{merged_event}
  & GRU training cost scales with N and sequence length; GPU recommended for large datasets \\
\bottomrule
\end{tabular}
\end{table}

Several limitations should be acknowledged. All empirical illustrations use a single item---the \textit{Party Invitations -- Can/Cannot Come} (U01a)---from the U.S. PIAAC sample. This choice was deliberate: maintaining one consistent example across all six methods allows direct comparison of methodological outputs within a controlled context, which is precisely the kind of cross-method comparability that motivated this paper. At the same time, the specific numerical results reported in the Supplementary Materials should be treated as illustrative rather than as benchmarks, because they reflect the particular demands of U01a and the characteristics of the U.S. sample. Certain methodological choices were also made for clarity rather than optimality; method-specific limitations and parameter sensitivities are summarized in Table~\ref{tab:method_comparison}, and researchers should tune these choices to the demands of their data and research question. Finally, the preprocessing pipeline targets PIAAC PS-TRE log files specifically; researchers working with other platforms such as PISA \citep{ray2003pisa} or NAEP \citep{mitchell1999grading} will need to adapt the core--ancillary pairing rules and event standardization procedures accordingly.

Future work may extend the present framework in several directions. Most directly, multi-item behavioral profiling would enable cross-item comparisons of problem-solving strategies, enable more systematic cross-method consistency checks, and support more precise ability estimation. The integration of timing and action content within unified models---for instance, combining reaction-time distributions with latent state structures---remains an open methodological challenge. Extension to other large-scale assessments such as PISA or NAEP would further test the generalizability of the preprocessing and analytical approaches introduced here.

\section*{Declarations}

\noindent\textbf{Funding.} Hwangbo and Park are co-first authors. This work was partially supported by the National Research Foundation of Korea [grant numbers NRF-2021S1A3A2A03088949 and RS-2024-00333701; Basic Science Research Program awarded to I.H.J.], the ICAN (ICT Challenge and Advanced Network of HRD) support program [grant number RS-2023-00259934], supervised by the IITP (Institute of Information \& Communications Technology Planning \& Evaluation), and the Ministry of Trade, Industry, and Energy (MOTIE), Korea, under the project ``Industrial Technology Infrastructure Program'' (RS-2024-00466693).

\noindent\textbf{Competing interests.} 
The authors have no competing interests to declare 
that are relevant to the content of this article.

\noindent\textbf{Ethics approval and consent to participate.}
Not applicable.

\noindent\textbf{Consent for publication.}
Not applicable.

\noindent\textbf{Data availability.} 
The PIAAC data are publicly available through the OECD 
(\url{https://www.oecd.org/en/data/datasets/piaac-1st-cycle-database.html}). 

\noindent\textbf{Materials availability.}
Not applicable.

\noindent\textbf{Code availability.}
All R code used in this study is publicly available at 
\url{https://github.com/hwangdabb/log-tutorial}.

\noindent\textbf{Authors' contributions.}
D.H. drafted the main manuscript, implemented the \path{R} 
code for the feature-based and model-based analyses, and 
reviewed the preprocessing sections. J.P. drafted the 
preprocessing sections, implemented the corresponding \path{Python} and \path{R} code, and reviewed the analytical sections. M.J. and I.H.J. conceived the study, provided methodological guidance throughout the project, and critically revised the manuscript. I.H.J. acquired funding. All authors read and approved the final manuscript.

\section*{Acknowledgments}

This work was partially supported by the National Research Foundation of Korea [grant number NRF-2021S1A3A2A03088949 and RS-2024-00333701; Basic Science Research Program awarded to I.H.J.] and the ICAN (ICT Challenge and Advanced Network of HRD) support program [grant number RS-2023-00259934], supervised by the IITP (Institute of Information \& Communications Technology Planning \& Evaluation), and the Ministry of Trade, Industry, and Energy (MOTIE), Korea, under the project ``Industrial Technology Infrastructure Program'' (RS-2024-00466693). Correspondence should be addressed to Ick Hoon Jin, Department of Applied Statistics, Department of Statistics and Data Science, Yonsei University, Seoul, Republic of Korea. E-Mail: ijin@yonsei.ac.kr. 

\newpage
\bibliographystyle{apalike}
\bibliography{reference}

\end{document}